\documentclass[11pt,preprint]{aastex}

\usepackage{verbatim}
\usepackage{natbib}
\usepackage{amsmath}
\usepackage{amsbsy}
\usepackage{lscape}
\usepackage{epstopdf}

\slugcomment{To appear in ApJ 720, 516 (2010)}
\shorttitle{Black hole--globular cluster relation}
\shortauthors{Burkert \& Tremaine}

\begin{document}

\title{A correlation between central supermassive black holes and the
  globular cluster systems of early-type galaxies} 

\author{Andreas Burkert\altaffilmark{1,2} and Scott Tremaine\altaffilmark{3}}

\altaffiltext{1}{University Observatory Munich, Scheinerstrasse 1,
D-81679 Munich, Germany} \altaffiltext{2}{Max-Planck-Fellow,
Max-Planck-Institute for Extraterrestrial Physics, Giessenbachstrasse
1, 85758 Garching, Germany} 

\altaffiltext{3}{School of Natural Sciences, Institute for Advanced
Study, Einstein Drive, Princeton, NJ 08540, USA}

\email{burkert@usm.lmu.de, tremaine@ias.edu}

\newcommand\msun{\rm M_{\odot}}
\newcommand\lsun{\rm L_{\odot}}
\newcommand\msunyr{\rm M_{\odot}\,yr^{-1}}
\newcommand\be{\begin{equation}}
\newcommand\en{\end{equation}}
\newcommand\cm{\rm cm}
\newcommand\kms{\rm{\, km \, s^{-1}}}
\newcommand\K{\rm K}
\newcommand\etal{{\rm et al}.\ }
\newcommand\sd{\partial}
\newcommand\Topspace{\rule{0pt}{2.6ex}}
\newcommand\Botspace{\rule[-1.2ex]{0pt}{0pt}}

\begin{abstract}
  Elliptical, lenticular, and early-type spiral galaxies show a
  remarkably tight power-law correlation between the mass $M_\bullet$
  of their central supermassive black hole (SMBH) and the number
  $N_{GC}$ of globular clusters: $M_\bullet = m_{\bullet/\star} \times
  N_{GC}^{1.08 \pm 0.04}$ with $m_{\bullet/\star} = 1.7\times
  10^5M_{\odot}$. Thus, to a good approximation the SMBH mass is the
  same as the total mass of the globular clusters.  Based on a limited
  sample of 13 galaxies, this relation appears to be a better
  predictor of SMBH mass (rms scatter 0.2 dex) than the
  $M_\bullet$--$\sigma$ relation between SMBH mass and velocity
  dispersion $\sigma$. The small scatter reflects the fact that
  galaxies with high globular cluster specific frequency $S_N$ tend to
  harbor SMBHs that are more massive than expected from the
  $M_\bullet$--$\sigma$ relation. 
\end{abstract}


\section{Introduction}

Supermassive black holes (SMBHs) have been detected in the centers of
many nearby galaxies \citep{kr95,mag98,gul09}. The SMBH masses are
correlated with several properties of their host galaxies
\citep{nov06}, in particular the velocity dispersion (the
$M_{\bullet}$--$\sigma$ relation, e.g.,
\citealt{fm00,geb00,tre02,gul09}) and the mass and luminosity of the
spheroidal component---the entire galaxy in the case of ellipticals or
the bulge in the case of lenticular and spiral galaxies
\citep{kor93,kr95,mh03,hr04}. \cite{sf09} find a tight correlation between
SMBH and dark matter halo masses. As the dark halo properties are inferred
from the number of globular clusters in a galaxy this
also indicates a connection between globular clusters and SMBHs.
These correlations suggest a strong link
between SMBH formation and galaxy formation, although the nature of
this link is poorly understood.

Numerous authors have investigated the possibility that the growth of
galaxies and their SMBHs is regulated by their interactions
\citep[e.g.,][]{hk00,bk01,som08,cat09}. SMBHs grow by several
mechanisms, including accretion of gas, swallowing stars whole, or
merging with other SMBHs acquired through a merger of their host
galaxies. The So\l tan argument \citep{sol82,yt02} suggests that gas
accretion is the dominant contributor to the SMBH mass budget, and
both observations \citep{san88} and simulations \citep{hop06} suggest
that much or most of this accretion occurs during mergers. SMBH growth
through gas accretion can release substantial amounts of energy that
can heat the interstellar gas, quench star formation, or even drive a
wind that sweeps the galaxy free of gas, thereby halting star
formation completely. Simulations of gas-rich galaxy mergers,
including seed black holes, can reproduce the observed
$M_{\bullet}$--$\sigma$ relation remarkably well given the simplicity
of the empirical prescriptions used to model the accretion and
feedback processes \citep{smh05,cox06,cro06,hop08,joh09a,joh09b}.

The origin of the seeds of SMBHs is controversial.  According to one
hypothesis, the seeds were remnants of the first generation of
metal-free, massive stars that formed at high redshifts. However, the
existence of bright quasars at redshifts of $z\gtrsim 6$ demonstrates
that SMBHs with masses exceeding $10^9M_\odot$ were already in place
less than a billion years after the Big Bang.  It is difficult for
black-hole remnants from first-generation stars to grow fast enough to
explain these observations \citep{may09}. A second hypothesis is that
the seeds are much larger (``intermediate-mass'') black holes of
$10^2$--$10^5M_\odot$, perhaps formed by the direct collapse of gas at
the centers of protogalaxies.

Globular clusters (GCs) are among the oldest stellar systems in the
universe and may have formed at the same time as the first stars.
Their high stellar densities, sometimes exceeding $10^5 M_\odot\hbox{
pc}^{-3}$, lead to a variety of complex dynamical phenomena
\citep{spi87,hh03,bt08}. Among these is mass segregation, through
which heavy, compact, stellar remnants---neutron stars and black
holes---spiral into the center by dynamical friction. Once 
these arrive at the center it is possible, though far from certain,
that they merge to form an intermediate-mass black hole
\citep{lee87,qs87,pz04,ku05}.  There is significant observational
evidence for intermediate-mass BHs with masses
$4\times10^3$--$4\times10^4M_\odot$ in the centers of several GCs
\citep{ger02,grh05,noy08,vdm10}, but this evidence is still
controversial\footnote{An additional uncertainty is whether some of
these systems might be tidally stripped dwarf galaxies masquerading as
GCs.}.  

The number of globular clusters in a galaxy, $N_{GC}$, is roughly
proportional to the total luminosity of the galaxy's spheroidal
component. This relation was quantified by \cite{hvdb81}, who
introduced the specific globular cluster frequency $S_N$, defined as
the number of GCs per unit absolute visual magnitude $M_V=-15$,
\begin{equation}
S_N \equiv N_{GC} \times 10^{0.4(M_V+15)}
\label{eq:sn}
\end{equation}
where $M_V$ is the magnitude of the spheroidal component. 

\cite{bs06} have summarized the progress that has been made in the quarter-century
since the work by Harris \& van den Bergh. It has become clear that star cluster
populations are powerful tracers of galaxy evolution and that the observed
correlations between globular cluster and galaxy properties provide valuable
information about their joint formation.  One of the most comprehensive
studies of early-type galaxies is by \cite{peng08}, who
measured specific frequencies for the globular cluster systems of 100
elliptical and lenticular galaxies in the Virgo cluster. They find
that early-type galaxies with intermediate luminosities ($-22 < M_V < -18$)
typically have $S_N \sim 1.5$, while luminous galaxies have $S_N\sim
2$--5. The dominant galaxy M87 has an even larger specific frequency \citep{rac68},
estimated by Peng et al.\ to be $S_N\simeq 13$. 

The formation of GCs is not well-understood (see, e.g., \citealt{bs06}
for a review). An important clue is that gas-rich merging galaxies
contain large numbers of young massive star clusters that presumably
formed in the merger \citep{sch87,ws95}. As this population of
clusters ages it is likely to evolve into a population of ``normal''
GCs \citep{fz01}. 
Another scenario is the combined formation of SMBH seeds and
globular clusters in super star-forming
clumps of gas-rich galactic disks at $z\sim2$ \citep{sgf10,mp96}.

In summary, (i) both the SMBH mass $M_\bullet$ and the total number of
GCs $N_{GC}$ are roughly proportional to the total luminosity of the
spheroidal component in early-type galaxies; (ii) GCs may provide the
black-hole seeds from which SMBHs grow; (iii) both the growth of SMBHs
and the formation of GCs appear to be associated with major
mergers or global gravitational instabilities in gas-rich protogalaxies. 
Given these observations, it is natural
to ask how the properties of the GC population in early-type galaxies
are correlated with the properties of their associated SMBHs. 

In this paper we show that there is a tight, power-law relation between the
mass of SMBHs and the total number of globular clusters in elliptical,
lenticular and early-type spiral galaxies. Remarkably, this relation
appears to have even less scatter than the classic relation between
SMBH mass and the velocity dispersion of the host galaxy. The relation
can be approximately characterized by the statement that the SMBH mass
equals the total mass in GCs.

\section{The BH-GC relation}

For this analysis we have selected all elliptical, lenticular and
early-type spiral galaxies with good estimates of the SMBH mass
$M_{\bullet}$ and the total number of globular clusters $N_{GC}$.

Most of the SMBH masses were taken from the recent compilation by
\cite{gul09}, with revised values for NGC 4486 (M87) and NGC 4649
(M60) from \cite{gt09} and \cite{sg10}, respectively. We also include
two unpublished SMBH measurements, provided by Karl Gebhardt: NGC 4472
(Shen, J., Gebhardt, K. et al. 2010, in preparation) and NGC 4594
(Gebhardt, K., Jardel, J. et al. 2010, in preparation).  Most of the
GC numbers were taken from the ACS Virgo Cluster Survey \citep[][first
choice]{peng08} or the compilation by \citet[][second
choice]{spi08}. Table 1 summarizes the data; the Table contains
entries for the Milky Way, the Sb spiral M31, and the dwarf elliptical
M32, but these are not included in the fits below. In three cases (NGC
1399, NGC 3379, and NGC 5128) where there are two good-quality values in
the literature that differ by two standard deviations or more, we have
included both estimates in our fits, each at half weight. Note that 
errors in GC numbers are significantly larger than Poisson, due 
to uncertainties in the extrapolation of the GC luminosity functions,
background subtraction, and corrections of the observed area of the 
galaxy to the whole system; the last of these is a particular concern because
the radial distribution of GCs does not always follow the radial distribution
of light. 

\begin{figure}[ht]
\begin{center}
\includegraphics[width=1.0\textwidth]{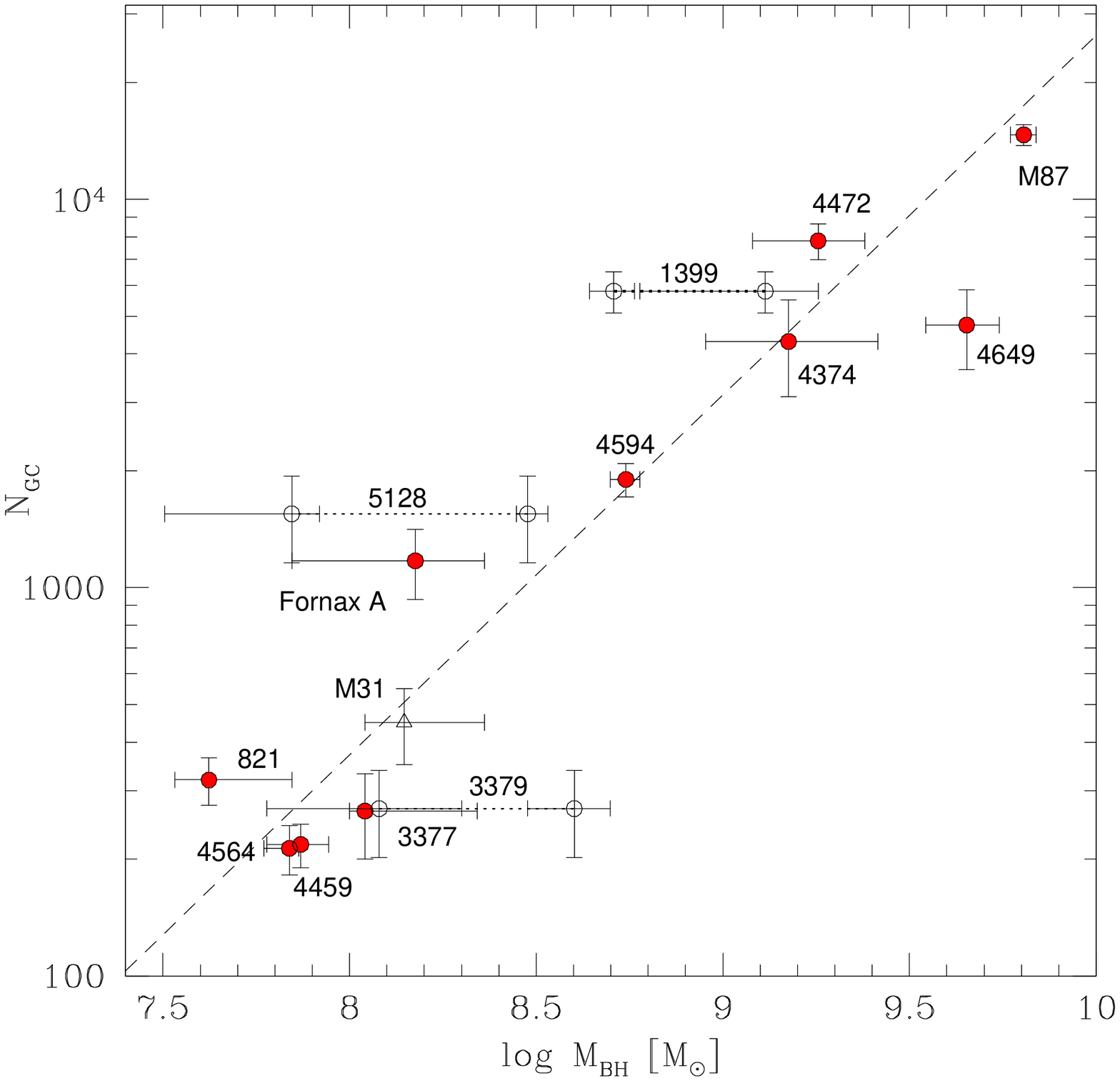}
\end{center}
\caption{
\label{fig:one} The number of globular clusters $N_{GC}$ is shown 
as a function of SMBH mass $M_{BH}$ for the 13 giant elliptical,
lenticular and early-type spiral galaxies in Table 1. Open circles
connected by dotted lines denote the galaxies NGC 1399, NGC 3379 and and NGC 5128
for which two estimates of the SMBH mass are given.  The dashed curve
shows the fit given by equation (\ref{eq:one}). The location of M31 is
also plotted as an open triangle, but this galaxy does not contribute
to the fit.}
\end{figure}

\begin{figure}[ht]
\begin{center}
\includegraphics[width=1.0\textwidth]{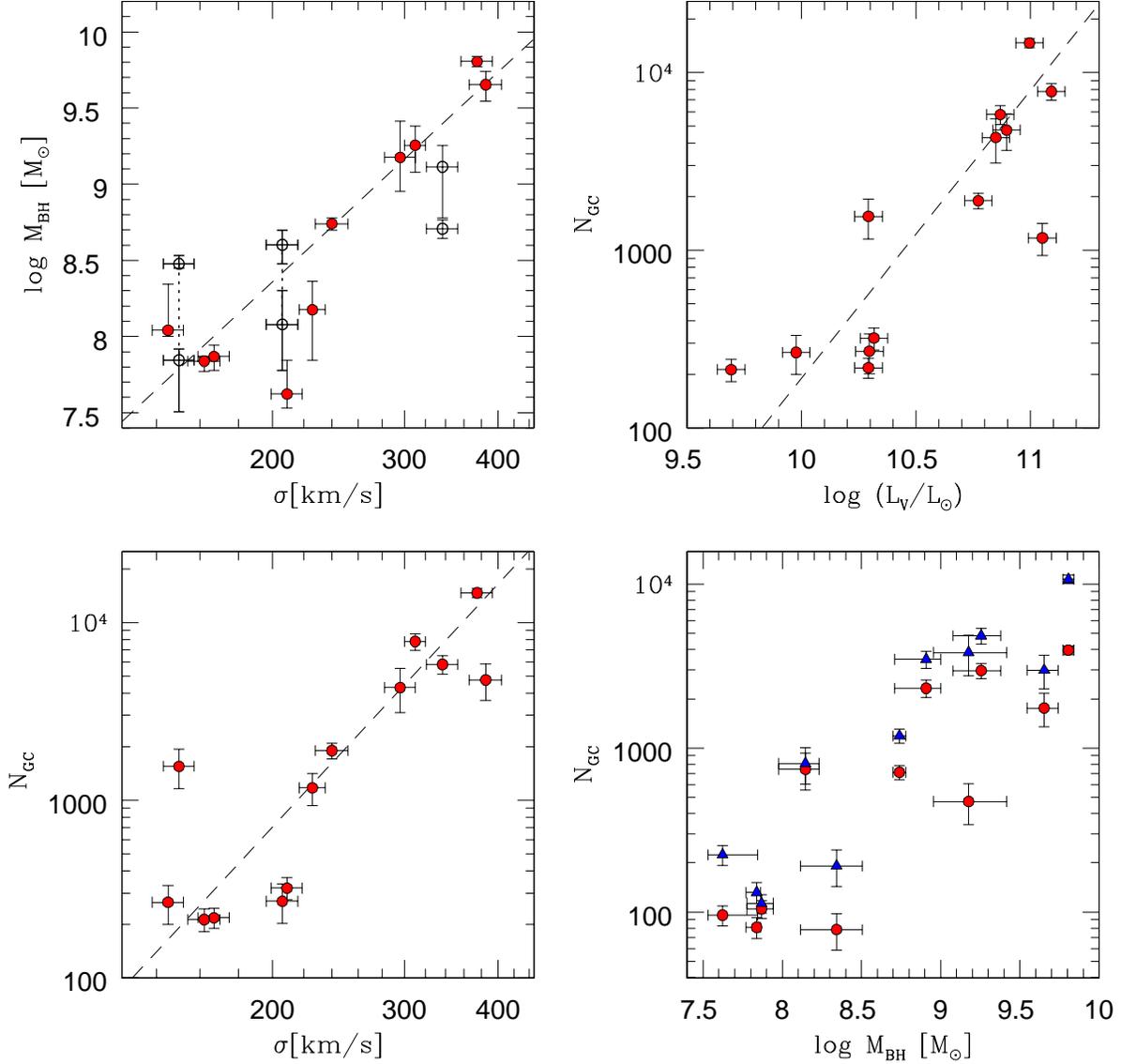}
\end{center}
\caption{
\label{fig:two}
The upper and lower left panels show the correlation between $M_{\bullet}$ or
$N_{GC}$, respectively, and the velocity dispersion. 
The dashed curve in the upper left panel corresponds to equation
(\ref{eq:msig}), and the line in the lower left panel is
$\log(N_{GC}/500)=0.15+4.54\log(\sigma/200\hbox{ km s}^{-1})$. 
In the upper right panel $N_{GC}$ is plotted versus the visual luminosity of
the host galaxy; the line shows the correlation given by equation (\ref{eq:ml}).
In the lower right panel blue triangles and red points correspond to the
number of blue and red globular clusters, respectively, versus SMBH mass; in
this panel the galaxies NGC 1399, NGC 3379 and NGC 5128 are represented by the geometric
mean of the two SMBH mass estimates.}
\end{figure}

The points in Figure \ref{fig:one} show SMBH mass $M_{\bullet}$ as a
function
of GC population $N_{GC}$. We fit this data to an assumed underlying
relation of the form $\log M_\bullet=\alpha + \beta \log(N_{GC})$ (all
logs in this paper are base 10). We determine the best-fit values of
$\alpha$ and $\beta$ by minimizing $\chi^2$ including errors in both
observational parameters, using the methods in \cite{tre02}.
There exists a surprisingly tight correlation (dashed line),
\begin{equation}
\log \frac{M_{\bullet}}{M_{\odot}} =  (8.14 \pm 0.04) + (1.08 \pm 0.04) \log
\frac{N_{GC}}{500};
\label{eq:one}
\end{equation}
the $\chi^2$ per degree of freedom is 6.6. For comparison, the
$M_{\bullet}$--$\sigma$ relation for the same sample, shown in the upper left
panel of Figure \ref{fig:two}, is
\begin{equation}
\log \frac{M_{\bullet}}{M_{\odot}} =  (8.36\pm0.04) + (4.57 \pm 0.25)
     \log \frac{\sigma}{200\hbox{ km s}^{-1}}
\label{eq:msig}
\end{equation}
with $\chi^2$ per degree of freedom of 8.5. Thus,
in this admittedly small sample ($N=13$), the correlation of SMBH mass
with globular-cluster number is actually {\em tighter} than the
classic correlation with velocity dispersion.

We have also carried out unweighted fits in which we ignore the observational
errors in dispersion, luminosity, and GC number and minimize the rms
residual $\epsilon$ in the log of the SMBH mass (weighted by the
observational errors in mass). For the mass-dispersion relation,
$\epsilon=0.30$ dex and for the mass-luminosity relation
$\epsilon=0.38$ dex, while for the mass vs.\ GC number relation
$\epsilon=0.21$ dex, substantially smaller than the other two. We used
the same procedure to fit to a relation of the form $\log
M_\bullet=\alpha + \beta_1\log\sigma + \beta_2\log L +\beta_3\log
N_{GC}$ and find that $\epsilon=0.19$, only marginally smaller than
the rms residual of 0.21 to the fit involving only $N_{GC}$ despite the
presence of the two additional free parameters $\beta_1$ and
$\beta_2$.

The smaller rms deviation $\epsilon$ seen in the correlation between
$M_\bullet$ and $N_{GC}$ implies that this is not a ``secondary''
correlation due to a tight correlation between $N_{GC}$ and bulge
properties, combined with a bulge--SMBH correlation.  Additional
evidence comes from the upper right panel of Figure \ref{fig:two},
showing the correlation of $N_{GC}$ with bulge visual luminosity $L_V$
which is less good ($\chi^2$ per degree of freedom of 34.8). We have also
checked the correlation of $N_{GC}$ with luminosity and dispersion using the
much larger sample of 62 galaxies in the sample of \cite{peng08} that also
have dispersions in HyperLeda and found fits of similar quality, with $\chi^2$
per degree of freedom of 23 and 27 respectively. 

\begin{figure}[ht]
\begin{center}
\includegraphics[width=1.0\textwidth]{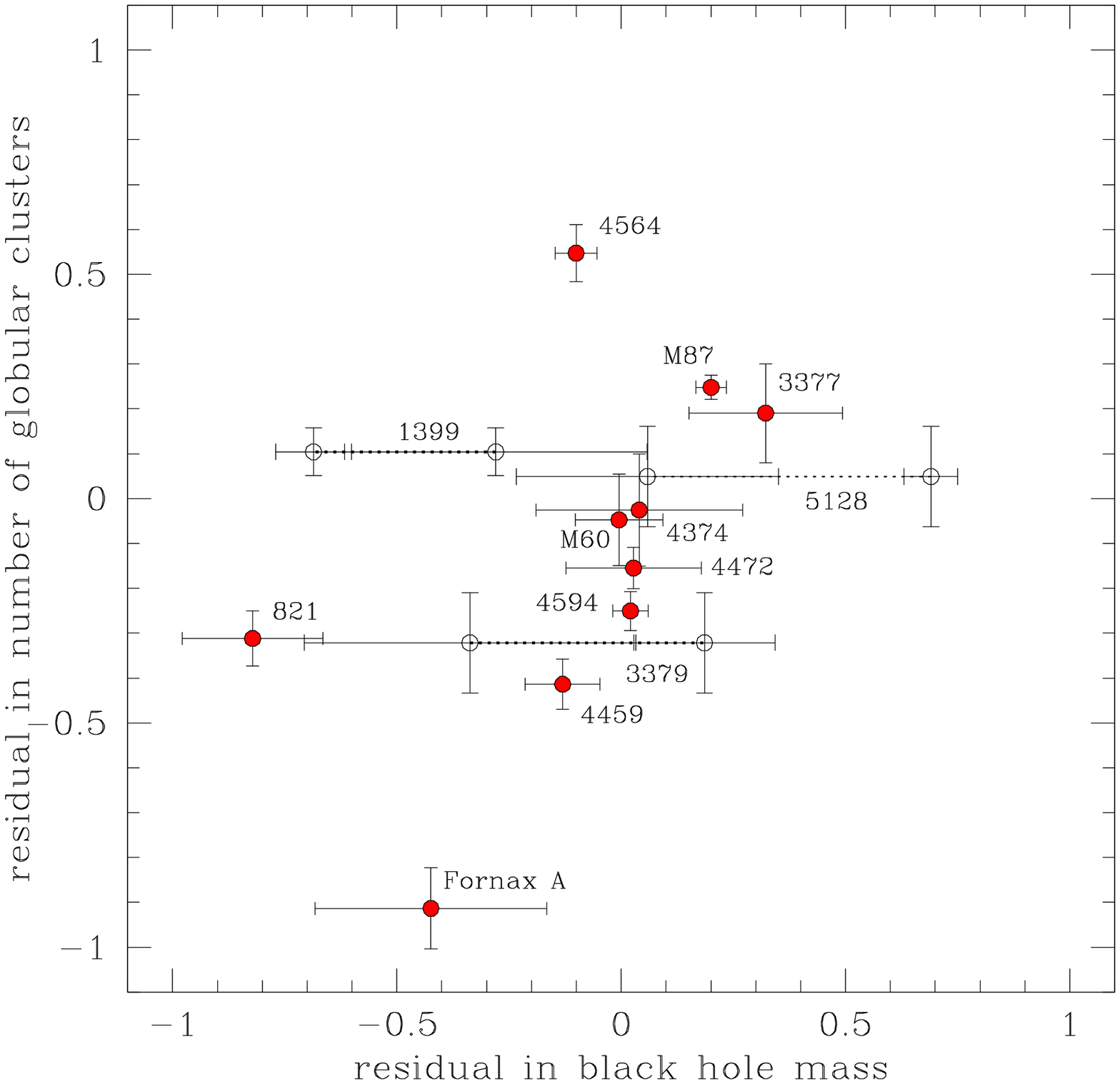}
\end{center}
\caption{
\label{fig:three}
Residuals from the best-fit relation between SMBH mass and velocity dispersion
(eq.\ \ref{eq:msig}) and the best-fit relation between number of globular
clusters and bulge luminosity (eq.\ \ref{eq:ml}). Dashed lines and open
circles denote the galaxies NGC 1399, NGC 3379 and NGC 1399 for which two estimates of
the SMBH mass are given.}
\end{figure}

Another look at the data is provided in Figure \ref{fig:three}, which
compares residuals from the best-fit $M_\bullet$--$\sigma$ relation
(eq.\ \ref{eq:msig}) on the horizontal axis to residuals from the
best-fit $N_{GC}$--$L$ relation
\begin{equation}
\log \frac{N_{GC}}{500} =  (-0.42\pm0.03) + 
(1.62\pm 0.04)\log \frac{L_V}{10^{10}L_\odot} 
\label{eq:ml}
\end{equation}
on the vertical axis. In general galaxies with positive (negative) residuals
in one quantity have positive (negative) residuals in the other.

It is helpful to look at a few individual galaxies.  Compare the galaxies
M87 (NGC 4486) and Fornax A (NGC 1316), which have similar
luminosities ($M_V=-22.7$ and $-22.8$ respectively). M87 has an SMBH
mass that is larger than expected for its dispersion by 0.20 dex,
while Fornax A contains an unusually small SMBH, by $-0.42$ dex. M87
has a specific frequency $S_N = 12.2$ that is a factor of 3 larger
than the average value for our sample of $S_N = 4.0$, while Fornax A
is characterized by an unusually small value, $S_N = 0.9$. A second
striking example is NGC 821, which has the largest residual from the
$M_\bullet$--$\sigma$ relation in our sample---$\log M_\bullet$ is
0.82 smaller than predicted. This galaxy also has a small specific
frequency, $S_N=1.3$, not far from the low value seen for NGC 1316. 

There is growing evidence that the masses of black holes in the most
luminous (core) galaxies have been underestimated in some cases:
including a dark halo \citep{gt09} or allowing for triaxiality
\citep{vdbdz} both tend to increase black-hole masses by a factor of
two or so in the few luminous core galaxies modeled so far. To test for the
effect of these revisions, we have increased the black-hole masses in
core galaxies, by a factor of four if the models accounted for neither
a dark halo nor triaxiality, and by a factor of two if the models
accounted for one of these two effects. This change slightly steepens
the $M_\bullet$--$N_{GC}$ relation---the best-fit slope in equation
(\ref{eq:one}) increases from $1.08\pm0.04$ to $1.17\pm0.04$---and
reduces the $\chi^2$ per degree of freedom from 6.6 to 4.8. It also
reduces the $\chi^2$ per degree of freedom for the
$M_\bullet$--$\sigma$ relation, from 8.5 to 5.0. With this revised
mass scale the unweighted best-fit $M_\bullet$--$N_{GC}$ and
$M_\bullet$--$\sigma$ give rms residuals of 0.20 and 0.26
respectively.

Assuming a mean globular-cluster mass $m_{GC}=2\times10^5M_\odot$, equation
(\ref{eq:one}) can be re-written in terms of the total mass of the
globular-cluster system $M_{GC}\equiv N_{GC}\,m_{GC}$,
\begin{equation}
\log \frac{M_{\bullet}}{10^8M_\odot} = (0.14\pm0.04) + (1.08\pm0.04)\log
\frac{M_{GC}}{10^8M_\odot}.
\label{eq:five}
\end{equation}
Remarkably, to a good approximation the mass of the SMBH is the {\em same} as
the total mass of the globular clusters.

\section{Local Group galaxies}

The compact elliptical galaxy M32 has a central SMBH with mass
$M_\bullet=3\times10^6M_\odot$ (Table 1) but no known globular
clusters. This exception to the correlation between $M_\bullet$ and
$N_{GC}$ observed in more luminous galaxies probably arises because
the globular clusters in M32 have spiraled into the center of the
galaxy through dynamical friction \citep{t76}.  The
characteristic inspiral time for an object of mass $m$ at initial
radius $r_i$ in a galaxy with dispersion $\sigma$ is \citep[][eq.\
8.12]{bt08}
\begin{equation}
  \tau=\frac{1.65}{\ln\Lambda}\frac{r_i^2\sigma}{Gm}=1.2\times10^{10}\,\hbox{yr}\frac{3}{\ln\Lambda}
\left(\frac{r_i}{0.5\,\hbox{kpc}}\right)^2\left(\frac{m}{2\times10^5\,\hbox{M}_{\odot}}\right)^{-1}
\frac{\sigma}{75\,\hbox{km s}^{-1}}
\end{equation}
where $\ln\Lambda$ is the standard Coulomb logarithm, $m$ is
normalized to the mean globular-cluster mass of $2 \times 10^5$
M$_{\odot}$ and the dispersion is normalized to M32's
\citep{cgj02}. According to this estimate GCs within 0.5 kpc would
have spiraled into the center within a Hubble time.  M32's effective
radius is only 0.14--0.18 kpc \citep{cgj02}, so it is not surprising
that most or all of the GCs in this galaxy have disappeared.

The nearby Sb spiral galaxy M31 has $450\pm100$ GCs (Table 1). According to
equation (\ref{eq:one}) this leads to a SMBH mass of $(1.2\pm0.4)\times
10^8M_{\odot}$, in excellent agreement with the observed value of
$1.4^{+0.9}_{-0.3} \times 10^8M_\odot$. Thus the $M_\bullet$--$N_{GC}$
correlation may work well for spiral galaxies as late as Hubble type Sb.  

The situation is less clear for the Milky Way, which contains a small
SMBH of mass $(4.3^{+0.4}_{-0.4})\times 10^6 M_\odot$ (Table
1). According to equation (\ref{eq:one}) this galaxy should contain
only about 20 GCs compared to the observed population of
$160\pm20$. The Milky Way is quite different from the elliptical,
lenticular, and early-type spiral galaxies that we have discussed so
far: it has a significantly later Hubble type (Sbc) and its bulge is
probably a pseudobulge \citep{bin09,shen10}. In fact, most of the GCs in the
Milky Way are associated with the Galactic halo, not the bulge, and as
SMBH mass appears to correlate with bulge luminosity rather than total
luminosity in spirals, it might be reasonable to expect that in
late-type spirals equation (\ref{eq:one}) applies only to the number
of bulge GCs. \cite{for01} argue that metal-rich GCs at radii $< 5$
kpc are mostly associated with the bulge, and estimate that there are
$35\pm4$ such clusters in the Milky Way. If we insert this number in
equation (\ref{eq:one}) we obtain an SMBH mass of
$(7.8\pm1.2)\times10^6M_\odot$; this is somewhat larger than the
observed mass but in reasonable agreement since some of the metal-rich
GCs in this region are likely to be disk GCs. We conclude that the
Milky Way offers no strong evidence for or against the proposed
correlation.

\section{Discussion and conclusions}

We have found that there is a strong correlation between the number of
globular clusters and the mass of the central SMBH in early-type
galaxies.  This correlation appears to be at least as tight as the
well known correlation between velocity dispersion and SMBH mass,
although this conclusion is based on only 13 galaxies. To a reasonably
good approximation, the BH-GC correlation simply says that the mass of
the central SMBH in an early-type galaxy is equal to the mass of its
GCs (eq.\ \ref{eq:five}). We suspect that the proportionality of the
SMBH mass to the total globular-cluster mass offers insight into their
formation processes, but the near-equality of the masses is a
coincidence.

Most galaxies have GC populations with a bimodal color distribution:
there are red (metal-rich) and blue (metal-poor) peaks, presumably
reflecting two sub-populations of GCs (e.g., \citealt{bs06}). It is
interesting to investigate whether the SMBH mass is correlated with
one or the other of these sub-populations. Table 1 shows the red
cluster fraction $f_{\rm red}$ for 11 galaxies, taken from
\cite{peng08} and \cite{rz04}. Note that $f_{\rm red}$ is rather
constant, with mean and standard deviation $0.3\pm0.1$. The lower
right panel of Figure \ref{fig:two} shows the $M_{\bullet}$--$N_{GC}$
correlation separately for the blue (triangles) and red (circles)
clusters. As expected from the small rms variation in the red cluster
fraction both correlations are of similar quality.

The origin of the $M_{\bullet}$--$N_{GC}$ relation is obscure. One
possibility is that both the growth of SMBHs and the formation of GCs
are associated with major mergers, so that galaxies that experienced a
recent major merger will have anomalously large SMBH masses and GC
populations. Another possibility is the correlated formation of
SMBH seeds and globular clusters in gas-rich young galaxies.

An important next step is to expand the sample of galaxies having both 
reliable SMBH masses and reliable globular-cluster populations.

\noindent
{\bf Acknowledgments:}
We thank Jeremiah Ostriker, Karl Gebhardt, Simon White and Leslie Sage for
interesting discussions and Karl Gebhardt for the use of unpublished data. We
also thank the referee, John Kormendy, for comments that substantially
improved the paper. 
The research of A.B. is supported by a Max Planck Fellowship and by the DFG
Cluster of Excellence ``Origin and Structure of the Universe''. S.T.'s
research is supported by NSF grant AST-0807432 and NASA grant NNX08AH24G.

\begin{landscape}
\begin{table}
\begin{center}
\begin{tabular}{ccccccccccc}
\hline
\hline
Galaxy& Type &    $M_{BH}$          &    $N_{GC}$    &   $\sigma$&$M_V$   &    $M_V$    & $S_N$     & $f_{\rm red}$ &   $B/T$ & Ref.\\ 
           &         &   ($M_{\odot})$      &                     &km           s$^{-1}$& (total)  &(sph.) &                 &        &                   & \\
\hline
  N821 \Topspace         & E4     & $4.2^{+2.8}_{-0.8} \times 10^7$ &  $320 \pm  45$  & $209\pm10$ &   $-21.0$   &   & $1.3 \pm 0.2 $  & 0.30 & 1 & 1,3 \\
  N1316=Fornax A         & E  & $1.5\pm0.8 \times 10^8$ &  $1173 \pm 240$ & $226\pm 9$ &   $-22.8$  &   & $0.9 \pm 0.2 $  & $-$ & 1 & 5,4 \\
  N1399$^{\rm a}$ & E1     &$1.3^{+0.5}_{-0.7} \times 10^9$  & $5800\pm 700$ & $337\pm16$   &   $-22.3$  &   & $7.0 \pm 0.8 $  &0.40  & 1 & 1,3 \\
                &        &$5.1\pm0.7 \times 10^8$  &
                &               &      &            &                &
                 &  &  \\
  N3377         & E5     & $1.1^{+1.1}_{-0.1} \times 10^8$ & $266  \pm66 $  & $145\pm7$   &  $-20.1$   &  & $2.4  \pm 0.6  $  &  $-$& 1  & 1,9 \\
  N3379         & E1     & $1.2^{+0.8}_{-0.6} \times 10^8$ & $270  \pm
  68$  & $206\pm10$   &  $-20.9$ &    & $1.2 \pm 0.3 $  & 0.29 & 1 &
  1,3,13 \\ 
        &      & $4\pm 1\times 10^8$ & 
   &   &   &    &   &  & &
  14 \\
  N4374         & E1     & $1.5^{+1.1}_{-0.6} \times 10^9$ & $4301 \pm 1201$& $296\pm14$   &  $-22.3$  &   & $5.2 \pm 1.4 $  & 0.11 & 1 &  1,2 \\
  N4459         & E2     & $7.4\pm1.4 \times 10^7$ & $218  \pm28$  & $167\pm 8$  &   $-20.9$  &   & $1.0 \pm 0.1 $  & 0.48  & 1 & 1,2 \\
  N4472         & E2     & $1.8\pm0.6 \times 10^9$ & $7813 \pm830$ & $310\pm10$   &   $-22.9$  &     & $5.4 \pm 0.6 $ & 0.38 & 1 & 6a,2,13 \\
  N4486=M87     & E1  & $6.4\pm0.5 \times 10^9$ & $14660\pm 891$& $375\pm18$   &   $-22.7$  &   & $12.2 \pm 0.7$ & 0.27 & 1 & 7,1,2 \\
  N4564         & S0      &$6.9^{+0.4}_{-1.0} \times 10^7$  & $213 \pm
  31$   & $162\pm8$    &  $-19.9$ & $-19.4$    & $3.7 \pm 0.5 $  & 0.38 & 0.63 & 1,2 \\
  N4594         & Sa     & $5.5\pm0.5 \times 10^8$ & $1900 \pm
  189$ & $240\pm12$   &   $-22.2$  &   $-22.1$ & $2.7\pm0.3$ & 0.37 & 0.9 & 6b,3,13 \\
  N4649         & E2     & $4.5\pm1.0 \times 10^9$ & $4745 \pm
  1099$& $385\pm19$  &   $-22.4$  &   & $5.2\pm 1.2 $  & 0.37  & 1 & 8,1,2 \\
  N5128=Cen A$^{\rm a}$ & E & $3.0^{+0.4}_{-0.2} \times 10^8$ &
  $1550 \pm 390  $& $150\pm7$   &   $-21.5$  &   & $3.9\pm 1.0$&  $0.48$
  & 1     & 1,3 \\
                &       & $7.0^{+1.3}_{-3.8} \times 10^7$  &                &             &                &               &      &  \\
\hline
  N224=M31 \Topspace         & Sb     & $1.4^{+0.9}_{-0.3} \times 10^8$ &  $450
  \pm 100$ & $160\pm8$ &   $-21.2$ & $-20.0$    & $4.5\pm 1.0 $    & $-$   &
  0.34    & 1,3,10 \\
  N221=M32          & dE2     & $3.1\pm0.6\times 10^6$  &  0
  & $75\pm3$      &$-16.8$ &         &        $0$   & $-$ & 1    & 1,14 \\
Milky Way       & SBbc    & $4.3\pm0.4\times 10^6$   & $160\pm20$
&$105\pm20$    & $-21.5$  &  note b &  note b &    $-$  &
note b      & 1,11,12 \\

\hline
\end{tabular}
\caption{\small Properties of galaxies with known SMBH masses and
  globular-cluster populations. The columns give morphological type,
  SMBH mass $M_{BH}$, number of globular clusters $N_{GC}$, velocity
  dispersion $\sigma$, absolute visual magnitude $M_V$ for the entire
  galaxy and for the spheroidal component alone as determined from
  $B/T$, globular cluster specific frequency $S_N$ relative to the
  spheroidal component (eq.\ \ref{eq:sn}), red cluster fraction
  $f_{\rm red}$ and bulge/total luminosity ratio $B/T$. We list the
  Local Group galaxies M31, M32, and the Milky Way below the
  horizontal line, but these are not included in our fits.
  References: (1) \cite{gul09}; (2) \cite{peng08}; (3) \cite{spi08};
  (4) \cite{gom01}; (5) \cite{now08}; (6a) Shen, J., Gebhardt, K. et
  al. 2010 (to be submitted); (6b) Gebhardt, K., Jardel, J. et
  al. 2010 (to be submitted); (7) \cite{gt09}; (8) \cite{sg10}; (9)
  \cite{kw01}; (10) \cite{ben05}; (11) \cite{gil09}; (12)
  \cite{for01}; (13) \cite{rz04}; (14) \cite{vdbdz}. $B/T$ ratios are
  taken from \cite{k09} for NGC 4564, \cite{gul09} for NGC 4594, and
  \cite{k10} for M31. $^{\rm a}$\cite{gul09} give two estimates for
  the SMBH masses in NGC 1399 and NGC 5128, and give half-weight to
  each in their fits. We adopt a similar procedure: we include both
  estimates but increase the error bars on each by $\surd{2}$ so that
  their combined contribution to the fit is the same as that of a
  single galaxy. $^{\rm b}$The Milky Way contains a pseudobulge.}
\end{center}
\end{table}
\end{landscape}

\end{document}